\documentclass[prd,showpacs,showkeys,floatfix,twocolumn,amsmath,amssymb,floatfix]{revtex4}
\usepackage{graphicx,color,dcolumn,booktabs,bm}
\usepackage{subfigure}
\usepackage{longtable,lscape}
\usepackage{amssymb}
\usepackage{indentfirst}
\usepackage{epsfig}
\usepackage{feynmf}   
\usepackage{epstopdf}   
\usepackage{slashed}  
\usepackage{cases}
\usepackage[pdfpages]{xcolor}
\definecolor{maroon}{RGB}{139,0,0}
\usepackage{multirow}
\usepackage{graphicx,color,dcolumn,booktabs,bm}
\usepackage[colorlinks, citecolor=blue,anchorcolor=red,menucolor=red, linkcolor=red,filecolor=red,runcolor=red,urlcolor=blue,frenchlinks=red, urlcolor=blue]{hyperref}
\def\pslash{p\!\!\!\slash }
\def\uslash{u\!\!\!\slash }

\def\Dslash{D\!\!\!\slash }

\begin{document}
\title{\color{maroon}{Effects of a dense medium on parameters of doubly heavy baryons}}
%
\author{K. Azizi$^{1,2,3}$}
\thanks{Corresponding author}
\author{N. Er$^4$}
\affiliation{
$^1$Department of Physics, University of Tehran, North Karegar Avenue, Tehran 14395-547, Iran\\
$^2$Department of Physics, Dogus University, Acibadem-Kadikoy, 34722 
Istanbul, Turkey \\
$^3$School of Particles and Accelerators, Institute for Research in Fundamental Sciences (IPM) P.O. Box 19395-5531, Tehran, Iran\\
$^4$Department of Physics, Abant \.{I}zzet Baysal University,
G\"olk\"oy Kamp\"us\"u, 14980 Bolu, Turkey
}
\begin{abstract}
The spectroscopic properties of the doubly heavy spin-$1/2$ baryons $\Xi_{QQ'}$, $\Xi'_{QQ'}$, $\Omega_{QQ'}$ and $\Omega'_{QQ'}$, with heavy quarks $Q$ and $Q'$ being $b$ or/and $c$, are investigated in cold nuclear matter. In particular, the behavior of the  mass of these particles with respect to the density of the medium in the range $\rho\in [0,1.4] ~\rho_{sat}$, with $\rho_{sat}=0.11^3 ~GeV^3$ being the  saturation density of nuclear matter, is investigated. From the shifts in the mass and vector self energy of the states under consideration, it is obtained that $\Xi_{QQ'}$ and $\Xi'_{QQ'}$ baryons with two heavy quarks and one $u$ or $d$  quark are affected by the medium, considerably.  It is also seen that the $\Omega_{QQ'}$ and $\Omega'_{QQ'}$ states, containing two heavy quarks and one $s$ quark do not see the dense medium, at all. The  value of mass for the $\Xi_{cc}$ state obtained at $\rho\rightarrow 0$ limit is nicely consistent with the experimental data. Our results on  parameters of  other members can be useful in the search for these states. The obtained results may also shed light on the future in-medium experiments aiming to search for the behavior of the doubly heavy baryons under extreme conditions.
\end{abstract}
%
%
%
\maketitle
\pagenumbering{arabic}
%
%
%
\section{Introduction}\label{sec:intro}
Although the doubly heavy baryons have been predicted by the quark model many decades ago  \cite{GELLMANN1964214,Zweig:1981pd,Zweig:1964jf}, only the spin$-1/2$ double-charmed baryon $\Xi_{cc}$ has been experimentally observed so far. Many models and approaches were used before the first observation of the $\Xi_{cc}$ state to understand the structure and spectrum of the doubly heavy baryons.  For instance, potential model and several versions of the bag model were used to calculate the mass spectrum of baryons with two charmed quarks surrounded by an ordinary or strange quark in Ref. \cite{10.1143/PTP.82.760}.  The SELEX Collaboration reported the first detection of $\Xi^+_{cc}$ state in the charge decay mode $\Xi^+_{cc} \rightarrow \Lambda_c^+ K^- \pi^+$ in 2002 \cite{Mattson_2002}. The measured mass for this state was $3519\pm 1$ MeV/$c^2$. Then in 2005, the same collaboration confirmed the same state in the charged decay mode $\Xi^+_{cc} \rightarrow p D^+ K^-$ \cite{OCHERASHVILI200518}. The updated measured mass,  $3518\pm 3$ MeV/$c^2$,  was in nice consistency with the previously measured value. The detection of this state by SELEX Collaboration triggered the theoretical studies devoted to the properties of the doubly heavy baryons using different approaches and models. For instance, using
double ratios of sum rules (DRSR), mass-splittings of doubly heavy baryons were obtained \cite{ALBUQUERQUE2010217}. Using  QCD sum rules, the doubly heavy baryon states were analyzed in Refs. \cite{ALIEV201259,Aliev_2013,Wang2018,PhysRevD.78.094007}.  The hypercentral constituent quark model (hCQM) was used in Refs. \cite{Shah2017,Shah2016} to obtain  the mass spectra
of doubly heavy baryons.  The mass spectra and radiative decays of doubly heavy baryons were investigated within the diquark picture in a relativized quark model in Ref. \cite{PhysRevD.96.114006}. In Ref. \cite{PhysRevD.97.054008}, an extended
chromomagnetic model  by further considering the effect of color interaction was used to study the mass spectra of all the lowest S-wave doubly and triply heavy-quark baryons.  In Refs. \cite{Li:2019ekr, Yu:2018com},  the  Bethe-Salpeter equation was applied for the mass spectra  of the doubly heavy baryons.  Discovery potentials of doubly heavy  baryons and their weak decays were analyzed for instance in Refs. \cite{Yu:2017zst,Shi:2019hbf,Shi:2019fph,Wang:2017azm}.

Among the results of theoretical studies, some of them were of great importance. These studies showed that the value of mass measured by SELEX Collaboration for the $\Xi_{cc}$ state remains considerably below the theoretical predictions. Thus, in Ref.  \cite{ALIEV201259}, the mass of this state was found as $3.72\pm 0.20$ GeV, which its central value remains roughly $200$ MeV above the experimental result. Motivated by these analyses the LHCb Collaboration  started to study this state. In 2017, this collaboration announced the observation of a   $\Xi^{++}_{cc}$ state in $ \Lambda_c^+   K^- \pi^+ \pi^-$ invariant mass, where the $\Lambda^+_c$ baryon was reconstructed in the decay mode $p K^- \pi^+$ \cite{PhysRevLett.119.112001}. The measured value for the mass of  $\Xi^{++}_{cc}$ state by LHCb Collaboration was $3621.40\pm72$ (stat.) $\pm ~0.27$ (syst.) $\pm~14(\Lambda_c^+)$ MeV/$c^2$, where the last uncertainty was due to the limited knowledge of the $\Lambda_c^+$ baryon mass. As is seen, the result of LHCb Collaboration differs considerably from the SELEX data. This tension was the starting points of a rush theoretical investigations deciding to explain the existing discrepancy between the SELEX and LHCb results. In Ref. \cite{Brodsky2018}, the authors showed that the intrinsic heavy-quark QCD mechanism for the hadroproduction of heavy hadrons at large  $x_F$
can resolve the apparent conflict between measurements of double-charm baryons by the SELEX fixed-target experiment and the LHCb experiment at the LHC collider.

 We hope that, by the development of experimental facilities, we will be able  to detect other members of the doubly heavy baryons. The production mechanism of doubly heavy baryons has an important place in the literature \cite{PhysRevD.49.555,KISELEV1994411,PhysRevD.54.3228,PhysRevD.57.4385,PhysRevD.64.034006,MA2003135,LI2007284,Zhong_Juan_2007,PhysRevD.83.034026,PhysRevD.86.054021,Martynenko2015,PhysRevD.90.094507,PhysRevD.93.114029,PhysRevD.95.074020,PhysRevD.98.113004,Brodsky2018,PhysRevD.98.094021,PhysRevD.97.074003}. Naturally, a doubly heavy baryon can be  produced using a two-step procedure: i-) in a hard interaction,  a double heavy diquark is produced perturbatively, ii-) and then it is transformed to the baryon within the soft hadronization process \cite{PhysRevD.98.113004}.   

Understanding the hadronic properties at finite temperature/density and  under extreme conditions are of great importance. Such investigations can help us  in the understanding of the natures and internal structures of the dense astrophysical objects like neutron stars as well as in analyzing  the results of the heavy ion collision and the in-medium experiments.  The spectroscopic parameters  of the light and single-heavy baryons   in medium  have been widely investigated (for instance see \cite{PhysRevD.94.114002,PhysRevLett.109.172001,PhysRevC.69.065210,DRUKAREV2003659,AZIZI2017147,Wang2011,Yasui:2018sxz,AZIZI2018422} and references therein). Although, the doubly heavy baryons have been widely studied in vacuum, the number of works devoted to the investigations of the properties of these baryons in a dense medium is very limited (for instance see Refs. \cite{Wang2012,PhysRevD.99.074012}). In Ref. \cite{PhysRevD.99.074012}  We investigated the fate of the doubly heavy spin-$3/2$  $\Xi^*_{QQ'}$ and $\Omega^*_{QQ'}$ baryons  in cold nuclear matter. The shifts on the physical parameters of these states due to nuclear medium were calculated at saturation medium density and compared with their vacuum values. In the present study, we investigate the  doubly heavy spin-$1/2$  $\Xi_{QQ'}$, $\Xi'_{QQ'}$, $\Omega_{QQ'}$ and $\Omega'_{QQ'}$  baryons in dense medium by the technique of the in-medium QCD sum rule. In particular, we discuss the behavior of different parameters related to the states under consideration with respect to the changes in the medium density in the range $\rho\in [0,1.4] ~\rho_{sat}$. We report the values of the masses and  vector self energies of the  spin-$1/2$ doubly heavy baryons at saturation nuclear matter density, $\rho_{sat}=0.11^3 ~GeV^3$, and compare the obtained results for the masses with their vacuum values in order to determine the order of shifts in the masses due to the dense medium.   The obtained results may shed light on the production and study of the in-medium properties of these baryons in future experiments.  Production of the doubly heavy baryons in dense medium requires simultaneous production of two pairs of the heavy quark-antiquark. A heavy quark from one pair, then,  needs to come  together with the heavy quark of the other pair, with the aim of forming a heavy diquark  with the total spin $1$ or $0$.  Meeting of the heavy diquark with a light quark forms a doubly heavy baryon in medium. These processes need that the quarks be in the vicinity of each other both in the ordinary and rapidity spaces. 

The rest of the paper is organized as follows. In next section we derive the in-medium QCD sum rules for the masses and vector self-energies of the doubly heavy spin-$1/2$ $\Xi_{QQ'}$, $\Xi'_{QQ'}$, $\Omega_{QQ'}$ and $\Omega'_{QQ'}$ baryons. In section III, using the input parameters, first, we fix the auxiliary parameters entering the sum rules by the requirements of the model. We discuss the behaviors of the physical quantities under consideration with respect to the changes in the density and calculate their values at saturation nuclear matter density. We compare the values of the masses obtained at $\rho\rightarrow 0$ limit with other theoretical predictions as well as the existing experimental result on the  doubly-charmed $\Xi_{cc}$ state. Section IV is devoted to the discussions and comments. We present  the in-medium light and heavy quarks propagators used in the calculations together with their ingredients: in-medium  quark, gluon and mixed condensates  in Appendix A. We reserve the Appendix  B to present the in-medium input parameters used in the numerical analyses.

%
%
\section{Sum rules for the in-medium parameters of the spin-$1/2$ doubly heavy baryons}

The aim of this section is to find the masses and vector self-energies of the doubly heavy spin-$1/2$ $\Xi_{QQ'}$, $\Xi'_{QQ'}$, $\Omega_{QQ'}$ and $\Omega'_{QQ'}$ baryons  in terms of QCD degrees of freedom as well as the auxiliary parameters entering the calculations. To this end, we employ the in-medium  QCD sum rule approach as one of the powerful and predictive non-perturbative methods in hadron physics. Here baryons without a prime refer to the symmetric states with respect to the exchange of two heavy-quark fields and those with a prime to the asymmetric states. For the classification of the ground state  spin-$1/2$ and  spin-$3/2$ baryons one can see for instance Ref.  \cite{PhysRevD.99.074012}. 

For the calculations of the physical parameters of the baryons under consideration the following in-medium correlation function is used:
\begin{equation}\label{corre}
\Pi^{S(A)}(p)=i\int{d^4 xe^{ip\cdot x}\langle\psi_0|\mathcal{T}[J^{S(A)}(x)\bar{J}^{S(A)}(0)]|\psi_0\rangle},
\end{equation}
where $p$ is the external four-momentum of the double heavy baryons, $|\psi_0\rangle$ is the ground state of the nuclear medium and $\mathcal{T}$ is the time ordering operator.  As we mentioned above, for the  doubly heavy spin-$1/2$ baryons, the interpolating currents can be symmetric ($J^{S}$) or anti-symmetric ($J^{A}$) with respect to the exchange of two heavy-quark fields. Considering the quantum numbers of the doubly heavy spin-$1/2$ baryons,  the symmetric and anti-symmetric interpolating currents can be written as
\begin{widetext}
\begin{eqnarray}
J^S &=& \frac{1}{\sqrt{2}} \epsilon_{abc}  \Bigg \{ \Big(Q^{aT}Cq^{b}\Big)\gamma_5 Q'^{c} +\Big(Q'^{aT}Cq^b\Big)\gamma_5Q^{c} +\beta \Big[ \Big(Q^{aT}C\gamma_5 q^b\Big)Q'^{c}  + \Big(Q'^{aT}C\gamma_5 q^b\Big)Q^{c} \Big]\Bigg \},\nonumber \\
 J^A &=&\frac{1}{\sqrt{6}} \epsilon_{abc} \Bigg \{ 2 \Big(Q^{aT}CQ'^{b} \Big)\gamma_5 q^c + \Big(Q^{aT}Cq^b \Big)\gamma_5Q'^{c}
 -  \Big(Q'^{aT}Cq^b \Big)\gamma_5Q^{c} + \beta \Big[ 2\Big(Q^{aT}C\gamma_5 Q'^{b} \Big)q^{c}  \nonumber \\
 &+&  \Big(Q^{aT}C\gamma_5  q^{b} \Big)Q'^{c} -   \Big(Q'^{aT}C\gamma_5  q^{b} \Big)Q^{c}  \Big]\Bigg \},
\end{eqnarray}
\end{widetext}
where $a, b$ and $c$ are color indices, $C$ is the charge conjugation operator, $\beta$ is an arbitrary mixing parameter and  $q$ is a light  quark field. In table~(\ref{table1}), we present the quark flavors of the doubly heavy spin$-1/2$ baryons.
\begin{table}[htp]
	\addtolength{\tabcolsep}{10pt}
	\begin{center}
\begin{tabular}{c|c|c|ccc}
	   \hline\hline
	  Baryon & $q$ &  $Q$	& $Q'$\\
	   \hline\hline
	  $\Xi_{QQ'}$ or $\Xi'_{QQ'}$ & $u$ or $d$ & $b$ or $c$ & $b$ or $c$ \\
	  $\Omega_{QQ'}$ or $\Omega'_{QQ'}$& $s$  & $b$ or $c$ & $b$ or $c$  \\
	   	   \hline\hline
\end{tabular}
\end{center}
\caption{The quark flavors of the doubly heavy spin$-1/2$ baryons.}
\label{table1}
\end{table}

As an example, let us briefly explain how the  current of the doubly heavy baryons in its anti-symmetric form is obtained. Considering the quark content and spin of these baryons the current $ J^A $ can be decomposed as
\begin{eqnarray}
\label{kazem}
 J^A &\sim & \epsilon^{abc} \Big\{ \Big(
Q^{aT} C \Gamma Q'^b \Big) \tilde{\Gamma} q^c + \Big(
Q^{aT} C \Gamma q^b \Big) \tilde{\Gamma} Q'^c 
\nonumber\\&-& \Big( Q'^{aT} C \Gamma q^b \Big) \tilde{\Gamma} Q^c  \Big\}~,
\end{eqnarray}
where $ \Gamma$, $ \tilde{\Gamma}=$ $ 1 $, $ \gamma_5 $, $ \gamma_\mu $, $ \gamma_5 \gamma_\mu$ or $ \sigma_{\mu\nu} $. Considering all  quantum numbers of the states under study, we shall determine 
$ \Gamma$ and $ \tilde{\Gamma}$. To this end, let us first consider the  transpose of the  quantity $ \epsilon^{abc} (Q^{aT} C \Gamma Q'^b )  $ from the first term in Eq. (\ref{kazem}):
\begin{eqnarray}
  [\epsilon^{abc} Q^{aT} C \Gamma Q'^b  ]^T&=&-\epsilon^{abc} Q'^{bT}  \Gamma^T C^{-1} Q^a\nonumber\\&=&\epsilon^{abc} Q'^{bT} C (C \Gamma^T C^{-1}) Q^a,
\end{eqnarray}
where  we used a simple theorem in the first line: If $ A=BD $, where $ A $, $ B $ and $ D $ are matrices whose elements are Grassmann numbers, then $ A^T=-D^TB^T $. In above equation, we also used  $ C^T=C^{-1} $ and $ C^2 $=-1.  The quantity, $ C \Gamma^T C^{-1} $ is equal to $ \Gamma $ for $\Gamma =1 $, $ \gamma_5 $ or $ \gamma_5 \gamma_\mu$ and it is equal to 
$ -\Gamma $ for $ \Gamma=\gamma_\mu $  or $ \sigma_{\mu\nu} $.  After switching the color dummy indices, one obtains 
\begin{eqnarray}
  [\epsilon^{abc} Q^{aT} C \Gamma Q'^b]^T  =-\epsilon^{abc} Q'^{aT} C  \Gamma  Q^b,
\end{eqnarray}
for  $\Gamma =1 $, $ \gamma_5 $ or $ \gamma_5 \gamma_\mu$ and 
\begin{eqnarray}
  [\epsilon^{abc} Q^{aT} C \Gamma Q'^b]^T  =\epsilon^{abc} Q'^{aT} C  \Gamma  Q^b,
\end{eqnarray}
for $ \Gamma=\gamma_\mu $  or $ \sigma_{\mu\nu} $. The right-hand side of last two equations are  anti-symmetric with respect to the replacement of two heavy quarks, $ Q\leftrightarrow Q' $. Using this property, we get
\begin{eqnarray}
  [\epsilon^{abc} Q^{aT} C \Gamma Q'^b]^T  =\epsilon^{abc} Q^{aT} C  \Gamma  Q'^b,
\end{eqnarray}
for  $\Gamma =1 $, $ \gamma_5 $ or $ \gamma_5 \gamma_\mu$ and 
\begin{eqnarray}
  [\epsilon^{abc} Q^{aT} C \Gamma Q'^b]^T  =-\epsilon^{abc} Q^{aT} C  \Gamma  Q'^b,
\end{eqnarray}
for $ \Gamma=\gamma_\mu $  or $ \sigma_{\mu\nu} $. From other side, the transpose of a $ 1\times 1 $ matrix $ \epsilon^{abc} Q^{aT} C \Gamma Q'^b $ should be equal to the same $ 1\times 1 $ matrix. Hence, we conclude that $\Gamma =1 $, $ \gamma_5 $ or $ \gamma_5 \gamma_\mu$.

The simplest way is to take the  baryons  interpolated by $J^A $ to have the same total spin and spin projection as the light quark $ q $. Therefore,  the spin of the diquark formed by two heavy quarks is zero. This implies  that $\Gamma =1 $ or $ \gamma_5 $. Thus, the two possible forms of the interpolating current $J^A $ can be written as
\begin{eqnarray}
j^A _{1} &= & \epsilon^{abc}   \Big(
Q^{aT} C  Q'^b \Big) \tilde{\Gamma}_1 q^c 
\nonumber\\ 
j^A _{2} &= & \epsilon^{abc}   \Big(
Q^{aT} C\gamma_5  Q'^b \Big) \tilde{\Gamma}_2 q^c.
\end{eqnarray}
The matrices $ \tilde{\Gamma}_1 $ and $  \tilde{\Gamma}_2$ are determined using the Lorentz and parity considerations. As $ j^A_{1} $ and $ j^A_{2} $ are Lorentz scalars, one should have $ \tilde{\Gamma}_1 $, $  \tilde{\Gamma}_2=$  $1 $ or $ \gamma_5 $. Finally,  considering the parity transformation leads to  $ \tilde{\Gamma}_1 =\gamma_5$ and $ \tilde{\Gamma}_2 =1$. Thus, 
\begin{eqnarray}
j^A_{1} &= & \epsilon^{abc}   \Big(
Q^{aT} C  Q'^b \Big) \gamma_5 q^c 
\nonumber\\ 
j^A_{2} &= & \epsilon^{abc}   \Big(
Q^{aT} C\gamma_5  Q'^b \Big)  q^c.
\end{eqnarray}
Obviously one uses their arbitrary linear combination, which leads to the  form 
\begin{eqnarray}
j^A &\sim & \epsilon^{abc} \Big[  \Big(
Q^{aT} C  Q'^b \Big) \gamma_5 q^c 
+ \beta   \Big(
Q^{aT} C\gamma_5  Q'^b \Big)  q^c\Big],\nonumber\\
\end{eqnarray}
for the first term in Eq. (\ref{kazem}), where we introduced the mixing parameter $ \beta $. Using similar arguments for the second and third terms in Eq. (\ref{kazem}),  we get the current $ J^A $ used in the calculations.  From a similar manner one can derive the symmetric current $ J^S $. We will use the symmetric current $ J^S $ and the anti-symmetric current $ J^A $ to interpolate the baryons without prime ( $\Xi_{QQ'}$,   $\Omega_{QQ'}$) and primed baryons ($\Xi'_{QQ'}$,   $\Omega'_{QQ'}$), respectively (for details see for instance Ref. \cite{ALIEV201259}).  We should also remark that the  currents  $ J^S $  and 
$ J^A $ are some possible currents which are obtained using the quantum numbers of the doubly heavy baryons under consideration. To construct the most 
general operators one may go through a similar procedure as explained in Ref. \cite{Chen:2008qv} for the light baryons.

The correlation function in Eq.~(\ref{corre}) can be calculated in two different ways:  in terms of hadronic parameters called phenomenological (or physical) side and in terms of QCD parameters like quark masses as well as in-medium quark, gluon and mixed condensates called QCD (or theoretical) side. By matching the coefficients of the selected  structures from both sides, one can get the sum rules for  different physical observables. The calculations are started in $ x $-space and then they are transferred to the momentum space. The Borel transformation is applied to both sides with the aim of suppressing the contributions of the higher states and continuum. As last step, a continuum subtraction procedure is applied with accompany of the quark-hadron duality assumption.

On the phenomenological side, the  correlation function is saturated with a complete set of the in-medium hadronic state carrying  the same quantum numbers as the related interpolating current. By performing the integral over four$-x$, we get
\begin{widetext} 
\begin{eqnarray}\label{phenom}
\Pi^{S(A)}(p)=-\frac{\langle\psi_0|J^{S(A)}(0)|B(p^*,s)\rangle\langle B(p^*,s)|\bar{J}^{S(A)}(0)|\psi_0\rangle}{p^{*2}-m^{*2}}  + ...,
\end{eqnarray}
\end{widetext} 
where dots are used to show the contributions of the higher states and continuum. The ket $|B(p^*,s)\rangle$ represents  the  doubly heavy spin-$1/2$ baryon state with spin $s$ and the in-medium momentum $p^*$.  Here $m^*$ is the modified  mass of the same state due to the dense medium. To proceed, the following  matrix elements  are defined:
\begin{eqnarray}\label{intcur}
\langle\psi_0|J^{S(A)}(0)|B(p^*,s)\rangle&=&\lambda^{*}u(p^*,s) \nonumber \\
\langle B(p^*,s)|\bar{J}^{S(A)}(0)|\psi_0\rangle&=&\bar{\lambda}^{*} \bar{u}(p^*,s)
\end{eqnarray}
where $\lambda^{*}$ is the modified coupling strength of the baryon  to nuclear medium and $u(p^*,s)$ is the in-medium Dirac spinor. After inserting Eq.~(\ref{intcur}) into Eq.~(\ref{phenom}) and performing summation over spins, we get the following expression for the phenomenological side of the correlation function:
\begin{eqnarray}\label{corre2}
\Pi^{S(A)}(p)&=&-\frac{\lambda^{*2}(\!\not\!{p^*}+m^{*})}{p^{*2}-m^{*2}} + ...  = -\frac{\lambda^{*2}}{\!\not\!p^{*}-m^{*}} + ... \nonumber \\
&=&-\frac{\lambda^{*2}}{(p^{\mu}-\Sigma_{\upsilon }^{\mu})\gamma_\mu-m^*}  + ...,
\end{eqnarray}
where $\Sigma_{\upsilon }^{\mu}$ is written in terms of the vector self energy the  doubly heavy spin-$1/2$ baryonic state  ($ \Sigma_{\upsilon} $) as: $\Sigma_{\upsilon }^{\mu}=\Sigma_{\upsilon} u^{\mu}+\Sigma'_{\upsilon} p^{\mu}$
with $u^{\mu}$ being the four-velocity of the nuclear medium and $\Sigma'_{\upsilon}$ is ignored because of its small value \cite{COHEN1995221}.
We shall work in the rest frame of the medium, $u^{\mu}=(1,0)$.

One can decompose the correlation function in terms of different structures as
\begin{eqnarray}\label{fullcorre}
\Pi^{S(A)}(p)&=&\Pi^{S(A)}_{\pslash}(p^2,p_0)\pslash+\Pi^{S(A)}_{\uslash}(p^2,p_0)\uslash \nonumber\\&+&\Pi^{S(A)}_{U}(p^2,p_0)U +  ..., \nonumber \\
\end{eqnarray}
where $U$ is the unit matrix and  $p_0$ is the energy of the quasi-particle. The coefficients of different  structures, i.e. the the invariant amplitudes $\Pi^{S(A)}_{i}(p^2,p_0)$ with $i = \pslash$,  $\uslash$ and  $U$  in above relation are obtained as
\begin{eqnarray}\label{BorStr}
\Pi^{S(A)}_{\pslash}(p^2,p_0)&=&-\lambda^{*2}\frac{1}{p^2-\mu^2},\nonumber \\
\Pi^{S(A)}_{\uslash}(p^2,p_0)&=&+\lambda^{*2}\frac{\Sigma_{\upsilon}}{p^2-\mu^2},
\nonumber \\
\Pi^{S(A)}_{U}(p^2,p_0)&=&-\lambda^{*2}\frac{m^*}{p^2-\mu^2} ,
\end{eqnarray}
where $\mu^2=m^{*2}-\Sigma^2_{\upsilon}+2p_{0}\Sigma_{\upsilon}$. After  applying the Borel transformation with respect to $p^2$, we obtain 
\begin{eqnarray}\label{3eqn6unknown}
\hat{B}\Pi^{S(A)}_{\pslash}(p^2,p_0)&=&\lambda^{*2}e^{-\mu^2/\mathcal{M}^2}, \nonumber \\
\hat{B}\Pi^{S(A)}_{\uslash}(p^2,p_0)&=&-\lambda^{*2}\Sigma_{\upsilon} e^{-\mu^2/\mathcal{M}^2},
\nonumber \\
\hat{B}\Pi^{S(A)}_{U}(p^2,p_0)&=&\lambda^{*2} m^* e^{-\mu^2/\mathcal{M}^2},
\end{eqnarray}
where $\mathcal{M}^2$ is the Borel mass parameter to be fixed later.

On the QCD side, we insert the explicit forms of the interpolating currents into the correlation function and contract the quark fields via the Wick theorem, as a result of which the following expressions for the symmetric and anti-symmetric parts are obtained in terms of the in-medium light quark ($S^{ij}_q$) and heavy quark ($S^{ij}_Q$)  propagators \cite{ALIEV201259}:
\begin{widetext}
\begin{eqnarray}\label{symmet}
\Pi^S (p) &=&i \kappa \epsilon_{abc}\epsilon_{a'b'c'} \int d^4 x e^{ip\cdot x} \Big\{ -\gamma_5 S^{cb'}_Q \tilde{S}^{ba'}_{q} S^{ac'}_{Q'} \gamma_5 - \gamma_5 S^{cb'}_{Q'} \tilde{S}^{ba'}_q S^{ac'}_{Q}\gamma_5 + \gamma_5 S^{cc'}_{Q'}\gamma_5 Tr \Big[ S^{ab'}_{Q} \tilde{S}^{ba'}_{q} \Big]\nonumber \\
&+&  \gamma_5 S^{cc'}_{Q}\gamma_5 Tr \Big[ S^{ab'}_{Q'} \tilde{S}^{ba'}_{q} \Big] +\beta \Big(-\gamma_5 S^{cb'}_Q \gamma_5 \tilde{S}^{ba'}_{q} S^{ac'}_{Q'} -\gamma_5 S^{cb'}_{Q'} \gamma_5 \tilde{S}^{ba'}_{q} S^{ac'}_{Q}  -  S^{cb'}_{Q}  \tilde{S}^{ba'}_{q} \gamma_5 S^{ac'} _{Q'} \gamma_5 \nonumber \\
&-& S^{cb'}_{Q'}  \tilde{S}^{ba'}_{q} \gamma_5 S^{ac'} _{Q} \gamma_5 + \gamma_5 S^{cc'}_{Q'} Tr \Big[ S^{ab'}_{Q} \gamma_5 \tilde{S}^{ba'}_{q} \Big] +S^{cc'}_{Q'}  \gamma_5  Tr \Big[ S^{ab'}_{Q} \tilde{S}^{ba'}_{q} \gamma_5 \Big] + \gamma_5 S^{cc'}_{Q} Tr \Big[ S^{ab'}_{Q'} \gamma_5 \tilde{S}^{ba'}_{q} \Big] \nonumber\\
&+&S^{cc'}_{Q}  \gamma_5  Tr \Big[ S^{ab'}_{Q'} \tilde{S}^{ba'}_{q} \gamma_5 \Big] \Big) + \beta^2 \Big(- S^{cb'}_Q \gamma_5 \tilde{S}^{ba'}_{q} \gamma_5 S^{ac'}_{Q'} - S^{cb'}_{Q'} \gamma_5 \tilde{S}^{ba'}_{q} \gamma_5 S^{ac'}_{Q} + S^{cc'}_{Q'}   Tr \Big[ S^{ba'}_{q} \gamma_5 \tilde{S}^{ab'}_{Q} \gamma_5 \Big] \nonumber \\
&+& S^{cc'}_{Q}   Tr \Big[ S^{ba'}_{q} \gamma_5 \tilde{S}^{ab'}_{Q'} \gamma_5 \Big]\Big)\Big\}_{\psi_0},
\end{eqnarray}
\begin{eqnarray}\label{asymmet}
\Pi^A (p) &=&\frac{i}{6} \epsilon_{abc}\epsilon_{a'b'c'} \int d^4 x e^{ip\cdot x} \Big\{2\gamma_5 S^{cb'}_Q \tilde{S}^{aa'}_{Q'} S^{bc'}_q \gamma_5 + \gamma_5 S^{cb'}_Q \tilde{S}^{ba'}_q S^{ac'}_{Q'}\gamma_5  -  2\gamma_5 S^{ca'}_{Q'} \tilde{S}^{ab'}_Q  S^{bc'}_q \gamma_5 \nonumber \\
&+& \gamma_5 S^{cb'}_{Q'} \tilde{S}^{ba'}_q  S^{ac'}_Q \gamma_5 -  2\gamma_5 S^{ca'}_{q} \tilde{S}^{ab'}_Q  S^{bc'}_{Q'} \gamma_5 + 2 \gamma_5 S^{ca'}_{q}   \tilde{S}^{bb'}_{Q'} S^{ac'}_Q \gamma_5  + 4 \gamma_5   S^{cc'}_{q} \gamma_5 Tr\Big[ S^{ab'}_Q\tilde{S}^{ba'}_{Q'}\Big] \nonumber\\
&+&  \gamma_5 S^{cc'}_{Q'}  \gamma_5 Tr\Big[ S^{ab'}_Q \tilde{S}^{ba'}_{q}\Big]  + \gamma_5 S^{cc'}_Q\gamma_5 Tr\Big[ S^{ab'}_{Q'} \tilde{S}^{ba'}_{q}\Big] + \beta\Big(2\gamma_5 S^{cb'}_Q  \gamma_5 \tilde{S}^{aa'}_{Q'} S^{bc'}_q  + \gamma_5 S^{cb'}_Q  \gamma_5 \tilde{S}^{ba'}_q S^{ac'}_{Q'} \nonumber \\
&-& 2\gamma_5 S^{ca'}_{Q'} \gamma_5 \tilde{S}^{ab'}_Q S^{bc'}_q  + \gamma_5 S^{cb'}_{Q'} \gamma_5 \tilde{S}^{ba'}_q S^{ac'}_Q - 2\gamma_5 S^{ca'}_q \gamma_5 \tilde{S}^{ab'}_Q S^{bc'}_{Q'}  + 2\gamma_5 S^{ca'}_q \gamma_5 \tilde{S}^{bb'}_{Q'}S^{ac'}_Q \nonumber \\
&+& 2 S^{cb'}_Q \tilde{S}^{aa'}_{Q'}  \gamma_5  S^{bc'}_q\gamma_5 + S^{cb'}_Q \tilde{S}^{ba'}_q \gamma_5  S^{ac'}_{Q'}\gamma_5 - 2 S^{ca'}_{Q'} \tilde{S}^{ab'}_Q  \gamma_5  S^{bc'}_q\gamma_5 + S^{cb'}_{Q'} \tilde{S}^{ba'}_q  \gamma_5  S^{ac'}_Q\gamma_5 \nonumber\\
&-& 2 S^{ca'}_q \tilde{S}^{ab'}_Q  \gamma_5  S^{bc'}_{Q'}\gamma_5 + 2 S^{ca'}_q \tilde{S}^{bb'}_{Q'}  \gamma_5  S^{ac'}_Q\gamma_5 + 4 \gamma_5 S^{cc'}_q Tr\Big[ S^{ab'}_Q \gamma_5 \tilde{S}^{ba'}_{Q'}\Big] + 4 S^{cc'}_q \gamma_5 Tr\Big[ S^{ab'}_Q \tilde{S}^{ba'}_{Q'}\gamma_5\Big] \nonumber \\
&+&  \gamma_5 S^{cc'}_{Q'} Tr\Big[ S^{ab'}_Q  \gamma_5 \tilde{S}^{ba'}_q \Big]  + S^{cc'}_{Q'} \gamma_5 Tr\Big[ S^{ab'}_Q \tilde{S}^{ba'}_q\gamma_5\Big] +  \gamma_5 S^{cc'}_{Q} Tr\Big[ S^{ab'}_{Q'}  \gamma_5 \tilde{S}^{ba'}_q \Big] + S^{cc'}_{Q} \gamma_5 Tr\Big[ S^{ab'}_{Q'} \tilde{S}^{ba'}_q\gamma_5\Big] \Big) \nonumber \\
&+& \beta^2 \Big(2 S^{cb'}_Q  \gamma_5 \tilde{S}^{aa'}_{Q'} \gamma_5 S^{bc'}_q  +S^{cb'}_Q  \gamma_5 \tilde{S}^{ba'}_q  \gamma_5  S^{ac'}_{Q'} -  2 S^{ca'}_{Q'}  \gamma_5 \tilde{S}^{ab'}_{Q} \gamma_5 S^{bc'}_q  + S^{cb'}_{Q'}  \gamma_5 \tilde{S}^{ba'}_{q} \gamma_5 S^{ac'}_Q \nonumber \\
&-&  2 S^{ca'}_{q}  \gamma_5 \tilde{S}^{ab'}_{Q} \gamma_5 S^{bc'}_{Q'}  + 2 S^{ca'}_{q}  \gamma_5 \tilde{S}^{bb'}_{Q'} \gamma_5 S^{ac'}_Q + 4S^{cc'}_q Tr\Big[ S^{ba'}_{Q'} \gamma_5 \tilde{S}^{ab'}_{Q} \gamma_5 \Big] + S^{cc'}_{Q'} Tr\Big[ S^{ba'}_{q} \gamma_5 \tilde{S}^{ab'}_{Q} \gamma_5 \Big]  \nonumber \\
&+& S^{cc'}_{Q} Tr\Big[ S^{ba'}_{q} \gamma_5 \tilde{S}^{ab'}_{Q'} \gamma_5 \Big] \Big)\Big\}_{\psi_0},
\end{eqnarray}
\end{widetext}
where $\tilde{S}=CS^T C$; and $\kappa=1$  for $Q=Q'$ cases and  $\kappa=\frac{1}{2}$ for the baryons with $Q\neq Q'$. The subindex $ \psi_0 $ represents that the calculations are done in a dense medium. The explicit expressions of the in-medium light and heavy quark propagators together with their ingredients including the in-medium quark, gluon and mixed condensates are presented in Appendix A. 

 On QCD side, the invariant amplitudes $\Pi^{S(A)}_{i}(p^2,p_0)$ corresponding  to different  structures in Eq.~(\ref{fullcorre}) can be represented as the following dispersion integral:
\begin{equation}\label{ }
\Pi^{S(A)}_{i}(p^2,p_0)=\int_{(m_Q+m_{Q'})^2}^{\infty} \frac{\rho^{S(A)}_i(s,p_0)}{s-p^2}ds,
\end{equation} 
 where $\rho^{S(A)}_i(s,p_0)$ are the corresponding two-point spectral densities,  which can be obtained from the imaginary parts of the correlation function. 
 In the QCD side,  the main goal is to calculate these spectral densities. To this end, we use the explicit forms of the in-medium light and heavy quarks propagators. By performing the integration over four$-x$, we transfer the calculations to the momentum space. But before that we use the following expression in order to   rearrange  the obtained expressions:
\begin{eqnarray}
\label{ }
\frac{1}{(x^2)^m}&=&\int \frac{d^Dk }{(2\pi)^D}e^{-ik \cdot x}i(-1)^{m+1}2^{D-2m}\pi^{D/2} \nonumber \\
&\times& \frac{\Gamma[D/2-m]}{\Gamma[m]}\Big(-\frac{1}{k^2}\Big)^{D/2-m},
\end{eqnarray}
which leads us to obtain  expressions with three four-dimensional integrals. It is very straightforward to perform the integral over four$-x$ leading to a Dirac delta function. The resultant Dirac delta function is used to perform the second four-integral. Finally,  the remaining four-integral is performed using the Feynman parametrization tool, which leads to the following equality as an example:
\begin{equation}
\label{ }
\int d^4 \ell\frac{(\ell^2)^m}{(\ell^2+\Delta)^n}=\frac{i\pi^2 (-1)^{m-n} \Gamma[m+2]\Gamma[n-m-2]}{\Gamma[2]\Gamma[n] (-\Delta)^{n-m-2}}.
\end{equation} 
To be able to obtain the imaginary parts corresponding to different structures, the following equality is applied:
\begin{equation}
\label{ }
\Gamma\Big[\frac{D}{2}-n\Big]\Big(-\frac{1}{\Delta}\Big)^{D/2-n}=\frac{(-1)^{n-1}}{(n-2)!}(-\Delta)^{n-2}ln[-\Delta],
\end{equation} 
where $ln[-\Delta]=i\pi+ln[\Delta]$ and the condition $\Delta>0$ brings constraints on the limits of the integrals over the Feynman parameters. As  examples, for the symmetric case of the correlation function, the spectral densities  $\rho^{S}_i(s,p_0)$ corresponding to  different structures are obtained as 
\begin{widetext}
\begin{eqnarray}
\rho^{S}_{\pslash}(s,p_0)&=&\frac{3}{32768 \xi^6 \pi^{12}} \int_0^1 dz \int_0^{1-z} dw \Bigg\{\xi^2 m_Q^4 w z (w+z) \Big[3 \big(\beta (5 \beta+2)+5\big) (w+z-1) (w+z)-2 \xi (\beta-1)^2\Big]\nonumber \\
&+&2 \xi m_Q^2 s w^2 z^2 (w + z - 1) \Big[\xi( \beta - 1)^2 -  4 \big( \beta (5  \beta + 2) + 5\big) (w + z - 1) (w + z)\Big] +  5 \Big[ \beta (5  \beta + 2) \nonumber \\
&+& 5\Big] s^2 w^3 z^3 (w + z - 1)^3 - \frac{\pi^2}{2}  \xi^2\langle \frac{\alpha_s}{\pi}G^2\rangle_{\rho} w z (w + z - 1) \Big[\big(\beta (3 \beta + 2) + 3\big) w^2 +  w \big(\beta (\beta (7 z - 3) \nonumber \\
&+& 6 z - 2) + 7 z - 3\big) + \big(\beta (3 \beta + 2) +  3\big) (z - 1) z\Big]\Bigg\}\Theta[L(s,z,w)] +  \frac{1}{384 \pi^{6}} \int_{z_1}^{z_2} dz \Bigg\{ \frac{9}{2}(\beta^2 - 1) m_Q  \langle \bar{u}u\rangle_{\rho} \nonumber \\
&+& (z-1) z \Big[4 (3 \beta^2+\beta +3) m_q \langle \bar{u}u\rangle_{\rho}-3 \big( \beta (3 \beta +2)+3\big) p_0  \langle u^{\dag} u\rangle_{\rho}- \big( \beta (3 \beta -2)+3\big)  \langle u^{\dag} iD_0u\rangle_{\rho}\Big]\Bigg\},\nonumber \\
\end{eqnarray}
\begin{eqnarray}
\rho^{S}_{\uslash}(s,p_0)&=& - \frac{1}{384 \pi^{6}} \int_{z_1}^{z_2} dz \Bigg\{ \frac{9}{2}  (\beta^2 - 1) m_q m_Q  \langle u^{\dag} u\rangle_{\rho}  - (z - 1) z \Big[3 \beta^2 p_0 \Big(m_q \langle \bar{u}u\rangle_{\rho} - 4   \langle u^{\dag} iD_0u\rangle_{\rho}\Big) \nonumber \\
&+& \beta \Big(3 \langle \bar{u}g_s\sigma G u\rangle_{\rho} - 2 m_q p_0  \langle \bar{u}u\rangle_{\rho} + 8 p_0  \langle u^{\dag} iD_0u\rangle_{\rho} -  6 s  \langle u^{\dag} u\rangle_{\rho}\Big) + 3 p_0 \Big(m_q  \langle \bar{u}u\rangle_{\rho} - 4  \langle u^{\dag} iD_0u\rangle_{\rho} \Big) \Big]\Bigg\},\nonumber\\
\end{eqnarray}
\begin{eqnarray}
\rho^{S}_U(s,p_0)&=&\frac{9(\beta^2-1)}{16384 \xi^4 \pi^{12}} \int_0^1 dz \int_0^{1-z} dw \Bigg\{\frac{2 m_Q s^2w^2z^2(w+z-1)^2(w+z)}{\xi}-3m_Q^3swz(w+z-1)(w+z)^2\nonumber \\
&+&\xi m_Q^5(w+z)^3 -\frac{\pi^2}{18} \langle \frac{\alpha_s}{\pi}G^2\rangle_{\rho} m_Q  \Big[z^2 \Big(\xi (8 w-1)+2 w^3\Big)+2 w z \Big(\xi (4 w-3)+(w-1) w^2\Big) \nonumber \\
&+&(w-1) w^2 (\xi-4 w^2)+z^3 \Big(\xi+2 (w-1) w\Big)+2 (w+2) z^4-4 z^5\Big]\Bigg\}\Theta[L(s,z,w)]  \nonumber \\
&+& \frac{1}{4096 \pi^{6}} \int_{z_1}^{z_2} dz \Bigg\{-16 (\beta^2-1)  m_Q  \Big[3 m_q \langle \bar{u}u\rangle_{\rho}-2 p_0  \langle u^{\dag} u\rangle_{\rho} \Big]+8 (\beta-1)^2 (z-1) z \Big[3\langle \bar{u}g_s\sigma G u\rangle_{\rho}  \nonumber \\
&+& 8 m_q p_0  \langle u^{\dag} u\rangle_{\rho}-6 s  \langle \bar{u}u\rangle_{\rho}\Big]+48   (\beta+1)^2 m_Q^2 \langle \bar{u}u\rangle_{\rho}+(\beta-1)^2 \langle \bar{u}g_s\sigma G u\rangle_{\rho} \Bigg\}, 
\end{eqnarray}
\end{widetext}
where $ \Theta[L(s,z,w)] $ is the unit-step function and 
\begin{eqnarray}
\label{ }
&&L(s,z,w)=\nonumber\\
&-&\frac{(w-1) \Big[m_Q^2 w \xi+z \Big(m_ {Q}^2\xi-s w (w+z-1)\Big)\Big]}{\xi^2},\nonumber \\
\xi&=&w^2+w (z-1)+(z-1) z,\nonumber \\
z_1&=& \frac{s-\sqrt{s^2-4 m_Q^2 s}}{2 s},\nonumber \\
z_2&=& \frac{s+\sqrt{s^2-4 m_Q^2 s}}{2 s}.
\end{eqnarray}

After applying the Borel transformation on the variable $p^2$ to the QCD side and  performing the continuum subtraction, we match the coefficients of different structures  from the physical as well as the QCD sides of the correlation function. As a result, we get the following in-medium sum rules  
\begin{eqnarray}\label{sumrules}
\lambda^{*2}e^{-\mu^2/\mathcal{M}^2}&=&\int_{(m_Q+m_{Q'})^2}^{s_0^*} ds \rho^{S(A)}_{\pslash}(s,p_0) e^{-s/\mathcal{M}^2}  , \nonumber \\
-\lambda^{*2}\Sigma_{\upsilon} e^{-\mu^2/\mathcal{M}^2}&=&\int_{(m_Q+m_{Q'})^2}^{s_0^*}  ds \rho^{S(A)}_{\uslash}(s,p_0) e^{-s/\mathcal{M}^2} , \nonumber \\
\nonumber \\
\lambda^{*2} m^* e^{-\mu^2/\mathcal{M}^2}&=&\int_{(m_Q+m_{Q'})^2}^{s_0^*} ds \rho^{S(A)}_{U}(s,p_0) e^{-s/\mathcal{M}^2} , \nonumber \\
\end{eqnarray}
where $s_0^*$ is the in-medium continuum threshold. By simultaneous solving of these coupled sum rules, we get the physical quantities in terms of the QCD degrees of freedom as well as the in-medium auxiliary parameters.

\section{Numercal Results}
In this section, we numerically analyze the sum rules obtained in previous  section in order to estimate the in-medium and vacuum mass as well as the vector self energy of the doubly heavy  spin-$1/2$,  $\Xi^{(')}_{QQ'}$ and $\Omega^{(')}_{QQ'}$ baryons. To this end, we need numerical values of input parameters like quark masses and in-medium as well as vacuum condensates including  quark, gluon and mixed condensates of different dimensions, whose values   are presented in Appendix B. 

The sum rules  for the physical quantities in Eq.~(\ref{sumrules}) contain three auxiliary parameters: the Borel mass parameter $\mathcal{M}^2$, the in-medium continuum threshold $s_0^*$ and mixing parameter $\beta$ entering the  symmetric and anti-symmetric spin-$1/2$ currents. We shall find their working regions  according to the standard prescriptions of the method such that the dependence of the physical quantities on these parameters are mild at these regions. To this end, we require the pole dominance as well as the convergence of the series of the operator product expansion. In technique language, the upper band of the Borel mass parameter is determined by requiring  that the pole contribution exceeds the contributions of the higher states and continuum, i. e,
 \begin{equation}
\label{ }
\frac{ \int_{(m_Q+m_{Q'})^2}^{s_0^*}~ds\rho_i^{S(A)}(s)e^{-s/M^2}}{\int_{(m_Q+m_{Q'})^2}^{\infty}~ds\rho_i^{S(A)}(s)e^{-s/M^2}} ~ > \frac{1}{2},
\end{equation}
while the lower limit of $\mathcal{M}^2$ is obtained demanding that the perturbative part exceeds the non-perturbative contributions and the series of non-perturbative operators converge. The continuum threshold is not totally arbitrary but it depends on the energies of the first excited states in the channels under consideration. We have not experimental information about the masses of the excited states under study yet. Hence, we consider the interval  $m_{QQ'}+E_1\leq\sqrt{s_0}\leq  m_{QQ'}+E_2$, where a energy from $ E_1 $ to $ E_2 $ is needed to excite the baryons, and  demand that the Borel curves are most flat and the pole dominance and the OPE convergence conditions are satisfied. Our analyses show that choosing the window  $m_{QQ'}+0.3~GeV\leq\sqrt{s_0}\leq  m_{QQ'}+0.5~GeV$ for  the doubly heavy baryons satisfies all these conditions. For the $\Xi_{cc}$ baryon, as an example, the mass in the limit $ \rho\rightarrow 0 $ ($ m_{\Xi_{cc}} $) shows a good stability with respect to $\mathcal{M}^2\in[3-5]$ GeV$^2$ in the interval $s_0^* \in[15.4-17.0] $ GeV$^2$, which is obtained from the above restrictions (see Fig. 1). From this figure, it is also clear that the variations of mass with respect to the continuum threshold are minimal in the chosen window.
\begin{figure}[hbt]
\label{QCDperx}
\includegraphics[width=0.48\textwidth]{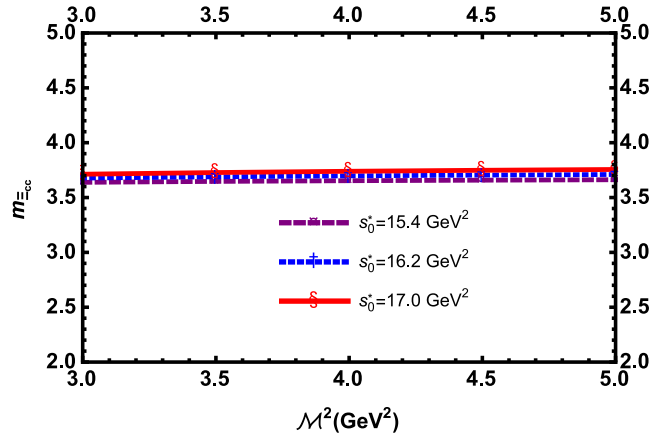}\\
\caption{$ m_{\Xi_{cc}} $ as a function of $\mathcal{M}^2$  at three different values of in-medium continuum threshold.}
\end{figure}
 For the  Borel mass parameter  and the in-medium continuum threshold the ranges presented  in table \ref{tab2} for different channels fulfill all the  requirements of the method. 
\begin{table}[!htbp]
	\addtolength{\tabcolsep}{10pt}
	\begin{center}\begin{tabular}{lcc}\hline \hline 
Channel &$\mathcal{M}^2$ (GeV$^2$) & $s_0^*$ (GeV$^2$)   \\
\hline\hline
$\Xi_{cc}$, $\Omega_{cc}$ & $[3-5]$ & $[15.4-17.0], [16.2-17.9]$   \\
$\Xi_{bc}$, $\Omega_{bc}$ & $[6-8]$ & $[49.3-52.1], [49.7-52.6]$   \\
$\Xi'_{bc}$, $\Omega'_{bc}$ & $[6-8]$ & $[50.3-53.1], [50.4-53.3]$   \\
$\Xi_{bb}$, $\Omega_{bb}$ & $[8-12]$ & $[105.2-109.4], [105.5-109.6]$   \\
\hline\hline
\end{tabular}
\end{center}
\caption{Working regions of the Borel mass $\mathcal{M}^2$ and the in-medium continuum threshold $s_0^*$ for different channels.}
\label{tab2}
\end{table}
 
 For determination of the reliable region of the auxiliary parameter $\beta$,  we plot the QCD side  of the result obtained using the structure $\pslash$ at $\Xi_{cc}$ channel, as an example,  as a function of $x$ in Fig. 2, where we use $x=\cos \theta$ with $\theta=\arctan\beta$ to explore the whole region $-\infty<\beta<\infty$ by sweeping the region $-1\leq x\leq1$. From this figure, we obtain the  following working intervals for $ x $, where the results are roughly independent of $ x $:
\begin{eqnarray}
 -1\leq x\leq-0.60 ~~ \textrm{and} ~~ 0.60\leq x\leq 1,
\end{eqnarray}
 for the vacuum and 
\begin{eqnarray}
 -1\leq x\leq -0.25 ~~ \textrm{and} ~~ 0.25\leq x\leq 1,
\end{eqnarray}
for the medium. Note that, the Ioffe current ($\beta =-1 $)  with $x=-0.71$ remains inside the reliable regions both for vacuum  and in-medium cases.   It is also clear that the medium enlarge the reliable regions of $\beta$, considerably. This is one of the main results of the present study. 
\begin{figure}[hbt]
\label{QCDperx}
\includegraphics[width=0.48\textwidth]{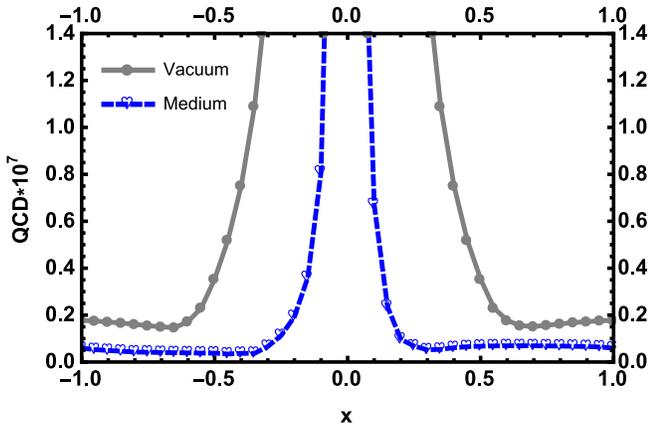}\\
\caption{Variation of QCD side as a function of $x$  obtained using the structure $\pslash$ at $\Xi_{cc}$ channel.}
\end{figure}

In order to check the pole contribution (PC), as an example for the $\Xi_{cc}$ channel and the structure $\pslash$, we plot PC as a function of $\mathcal{M}^2$ at three fixed values of the in-medium continuum threshold and at saturation nuclear matter density and $x=0.85$ in Fig. 3. From this figure we obtain, in average,  PC $=70\%$ and PC $=49\%$ at lower and higher limits of the Borel parameter, respectively. Our analyses  show also that, with the above working windows for  the auxiliary parameters, the series of sum rules converge, nicely.
\begin{figure}\label{PoleCont}
\centering
\includegraphics[width=0.48\textwidth]{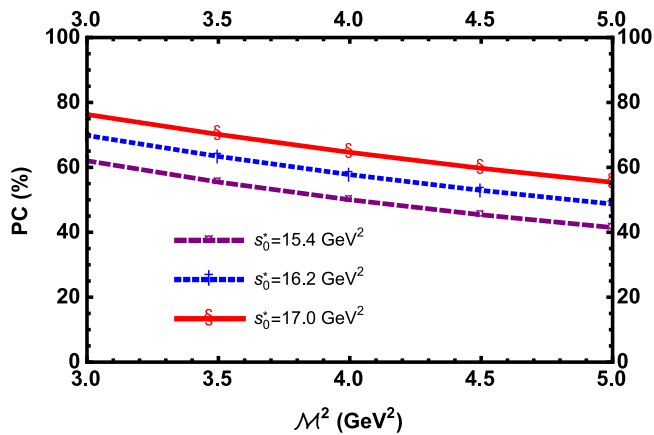}
\caption{ PC with respect to $\mathcal{M}^2$ for the $\Xi_{cc}$ channel and the structure $\pslash$ at $\rho=\rho_{sat}$ and $x=0.85$.}  
\end{figure}

We plot the ratio of the in-medium mass to vacuum mass, i.e. $m^*/m$, with respect to  $\mathcal{M}^2$ for the doubly heavy  $\Xi_{QQ'}$ and $\Omega_{QQ'}$ baryons at average value of the continuum threshold and at the saturation nuclear matter density in Fig. 4.  This figure shows that in the selected windows for $\mathcal{M}^2$, $m^*/m$ for all members show good stability against the variations of   $\mathcal{M}^2$. It is also clear that the $\Omega_{QQ'}$ baryons are not affected by the medium at the saturation medium density, while the mass of $\Xi_{QQ'}$ baryons reduce to nearly  $80\%$ of their vacuum values at  saturation nuclear matter density. Note that the vacuum masses  are obtained from the in-medium calculations in the limit $\rho \rightarrow 0$.
\begin{figure}[hbt]
\label{Borel}
\includegraphics[width=0.48\textwidth]{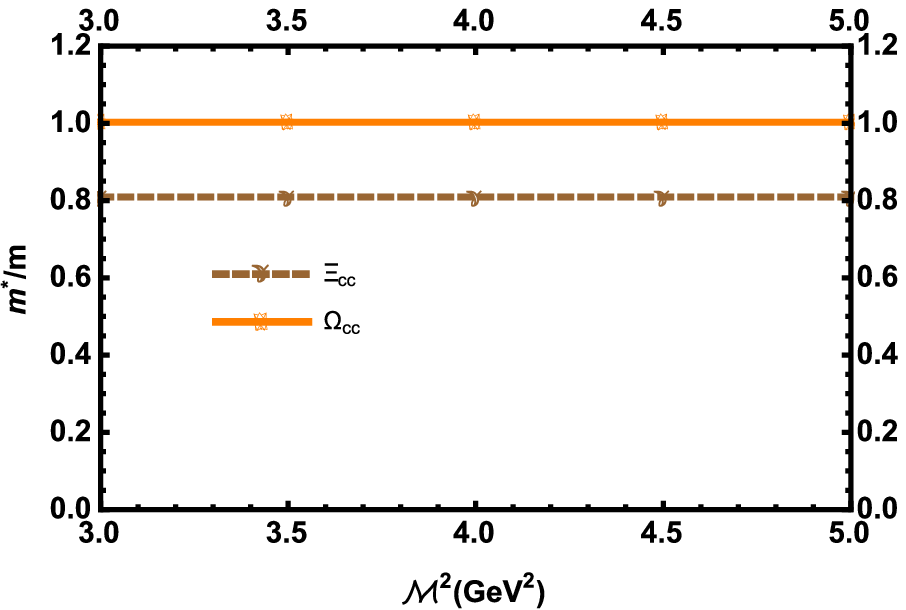}\\
\includegraphics[width=0.48\textwidth]{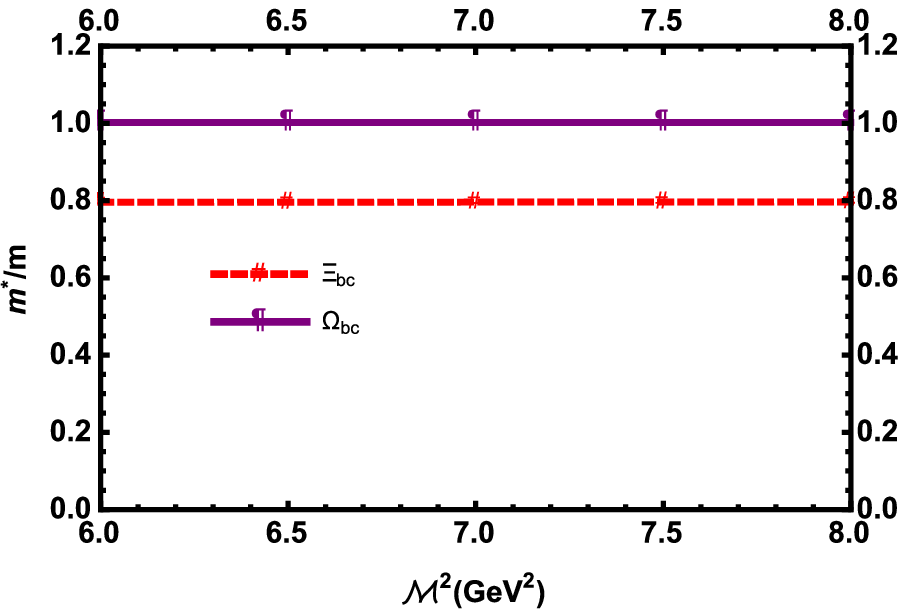}\\
\includegraphics[width=0.48\textwidth]{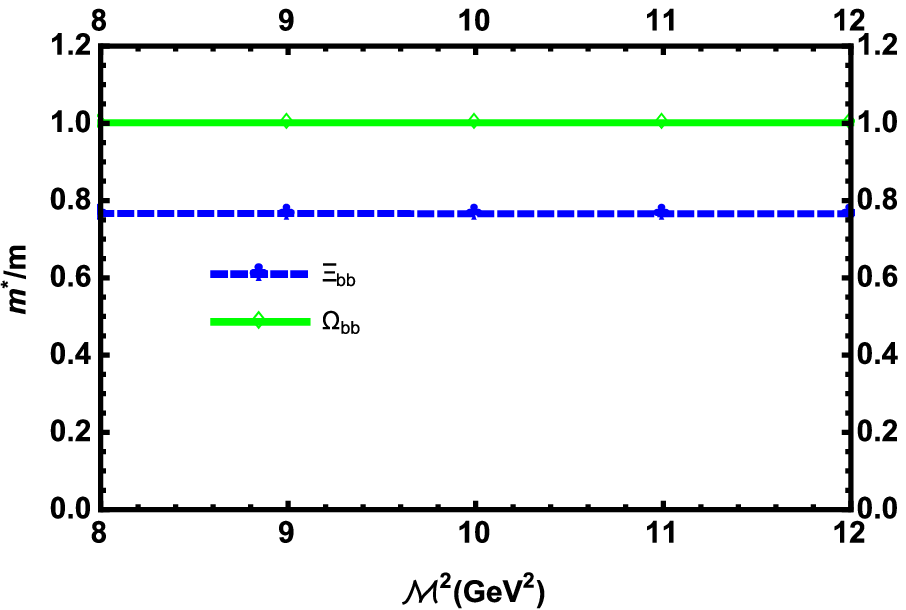}\\
\caption{The in-medium mass to vacuum mass ratio $m^*/m$ with respect to Borel mass $\mathcal{M}^2$ for the spin-$1/2$  doubly heavy  $\Xi_{QQ'}$ and $\Omega_{QQ'}$ baryons at average value of continuum threshold and at  saturation nuclear matter density.}
\end{figure}
\begin{figure}[hbt]
\label{Density}
\includegraphics[width=0.48\textwidth]{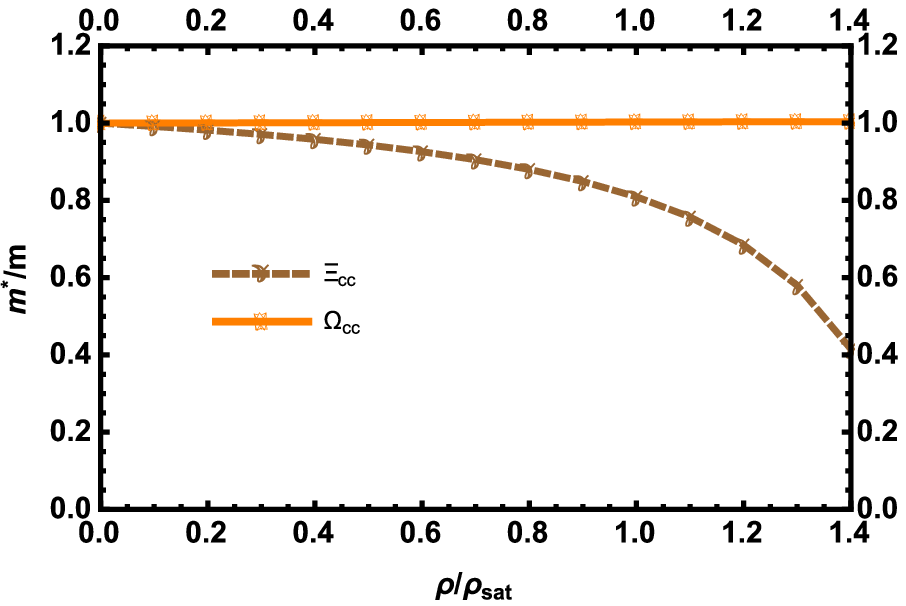}\\
\includegraphics[width=0.48\textwidth]{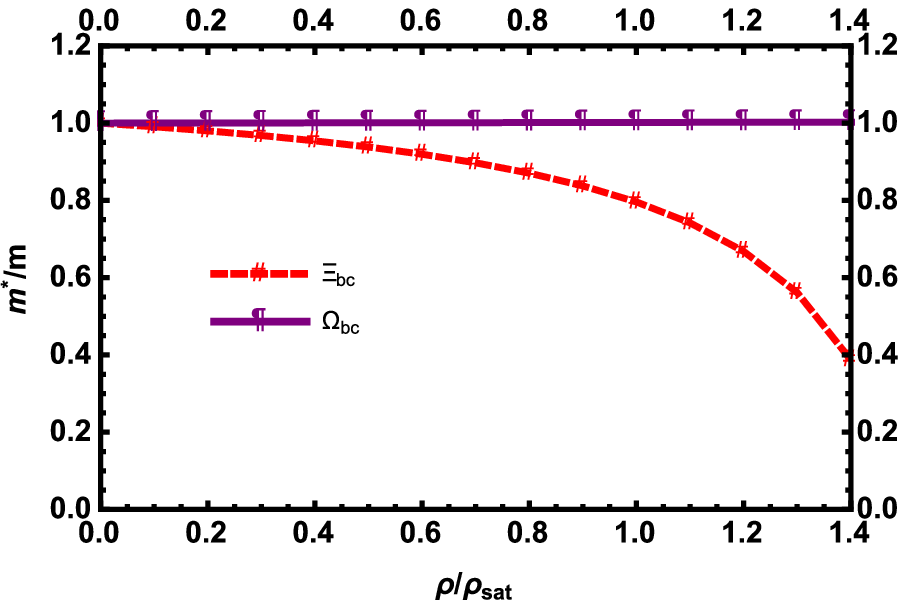}\\
\includegraphics[width=0.48\textwidth]{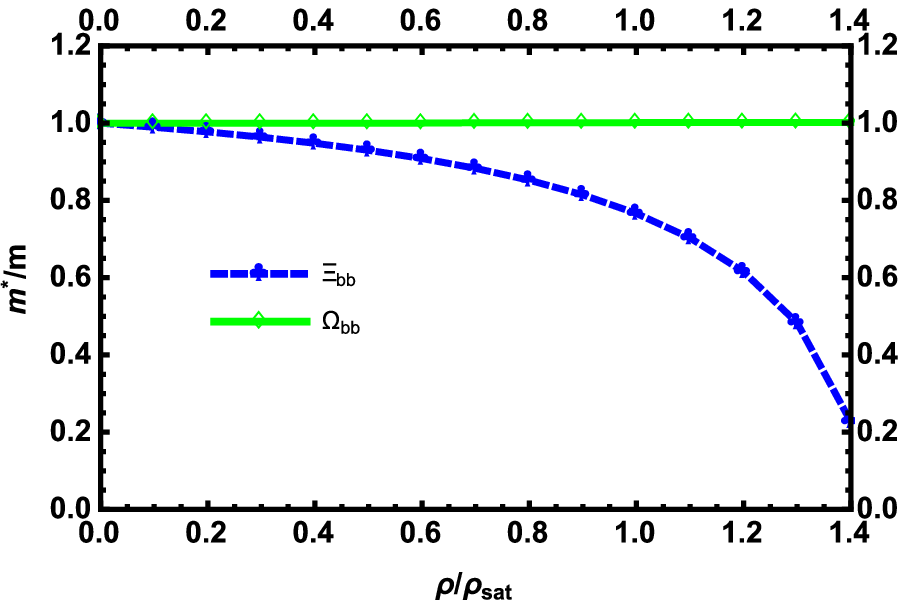}\\
\caption{The in-medium mass to vacuum mass ratio $m^*/m$ of the doubly heavy  $\Xi_{QQ'}$ and $\Omega_{QQ'}$ baryons  with respect to $\rho/\rho_{sat}$ at average values of continuum threshold and Borel mass parameter.}
\end{figure}

The main goal of the present study is to investigate the behavior of the mass of the states under consideration with respect to the density of the medium. In this accordance, in Fig. 5, we depict the ratio  $m^*/m$ with respect to $\rho/\rho_{sat}$ for the doubly heavy  $\Xi_{QQ'}$ and $\Omega_{QQ'}$ baryons at average values of the continuum threshold, Borel mass parameter and considering the reliable regions of the mixing parameter $\beta  $. We consider the range $ \rho\in [0, 1.4] \rho_{sat}$ to investigate the behavior of the masses, where the previously presented value of $\rho_{sat}  $  is equivalent to roughly $ 1/5 $ of the density of the neutron stars' core. From this figure we read that the $\Omega_{QQ'}$  baryons  containing two heavy and one strange quarks do not see the dense medium at all.  Similar behavior is the case for the doubly heavy $\Omega'_{QQ'}$ baryons.  The doubly heavy baryons $\Xi^{(')}_{QQ'}$ with the quark contents of two heavy quarks and one up or down quark, however,   are affected by the medium, considerably. Such that, as it is seen from Fig. 5, the mass of the  baryons  $\Xi_{cc}, \Xi_{bc}$ and $\Xi_{bb}$ reach to $42\%$, $40\%$ and $24\%$ of their vacuum values at $\rho/\rho_{sat}=1.4$, respectively.     At  $\rho/\rho_{sat}=1$, the in-medium mass to vacuum mass ratios for these baryons are obtained as $0.81$,  $0.79$  and $0.77$, respectively. The negative shifts on the masses due to the medium show that these baryons are attracted by the medium, considerably.

The saturation nuclear matter density is an important point that we would like to present the numerical values of the modified masses as well as the vector self-energies at all channels under study. To this end,  in table  \ref{tab3}, we collect the average values of these quantities at $  \rho=\rho_{sat}$ together with the vacuum masses of the  doubly heavy  $\Xi^{(')}_{QQ'}$ and $\Omega^{(')}_{QQ'}$ baryons  obtained in the limit $ \rho\rightarrow 0$. The uncertainties in the numerical results are due to the errors in the values of the input parameters as well as the uncertainties in determination  of the  working windows of the  auxiliary parameters. It would be instructive to check the impact of some important input parameters like  $\sigma_{\pi N}$ and its strange
counterpart $\sigma_{s N}$ on the finite density behavior of the studied hadrons. They appear as $y=\frac{2m_q}{m_s}\frac{\sigma_{sN}}{\sigma_{\pi N}}$ in the calculations. As we present in the Appendix B, we use  the average of  values obtained in Refs. \cite{PhysRevD.87.074503} and \cite{Dinter:2011za} for this parameter.  However,  different methods and approaches obtain different values for $ y $. In  Ref. \cite{GUBLER20191} the numerical values of  $\sigma_{sN}$ and  $\sigma_{\pi N}$  are collected from different sources \cite{PhysRevLett.116.172001,PhysRevD.94.054503,PhysRevLett.116.252001,PhysRevD.93.094504,PhysRevD.87.074503,PhysRevD.98.054516,SEMKE2012242,PhysRevD.91.051502}, which give $ y $ in the interval $[-0.05, 0.36]$ considering the corresponding errors. Taking into account this interval we see that  the mass of, as an example $\Omega_{cc}$ state containing a strange quark, is changed maximally by    $0.09\%$  compared to the value considered in the Appendix B. Therefore, the  effect of $ y $ on the parameters of the doubly heavy baryons is very weak. Our analyses show that the auxiliary parameters are sources of the main uncertainties in the presented results. 

 By comparison of the vacuum masses with the masses at saturation point, we see that $\Omega^{(')}_{QQ'}$ baryons are not aware of the environment. The negative shifts on the masses of the $\Xi^{(')}_{QQ'}$ baryons, however, refer to the strong scalar attractions of these states by the dense medium.  The baryons $\Omega^{(')}_{QQ'}$ gain small vector self-energies in dense medium compared to the $\Xi^{(')}_{QQ'}$ baryons that receive large positive vector self-energies referring to the strong vector repulsion of these states by the nuclear medium. 
\begin{table}[h]
	\addtolength{\tabcolsep}{10pt}
	\begin{center}\begin{tabular}{lccc}\hline \hline 
&$m(\rho=0)$ & $m^*(\rho=\rho_{sat})$  & $\Sigma_{\upsilon} (\rho=\rho_{sat})$  \\
\hline\hline
$\Xi_{cc}$  & $3.65 \pm 0.05$ & $2.96 \pm 0.04$ & $0.71\pm 0.10$ \\
$\Omega_{cc}$   & $3.66 \pm 0.09$ & $3.66 \pm 0.09$ & $ 0.12\pm 0.03$ \\
$\Xi_{bc}$  & $6.49 \pm 0.04$ &  $5.17 \pm 0.03$ &$1.07 \pm  0.15$\\
$\Omega_{bc}$   & $6.51 \pm 0.05$ & $6.51 \pm 0.05$ &  $0.14 \pm 0.04$ \\
$\Xi_{bb}$  & $9.17 \pm 0.06$ & $6.97 \pm 0.05$ & $1.45 \pm 0.20$ \\
$\Omega_{bb}$   &  $9.20 \pm 0.06$ & $9.19 \pm 0.06$ & $0.14 \pm 0.04$ \\
$\Xi'_{bc}$  & $6.53 \pm 0.09$&  $5.01 \pm 0.07$ & $1.05 \pm 0.14$\\
$\Omega'_{bc}$   &$7.01 \pm 0.04$ &$6.98 \pm 0.04$ &   $0.13 \pm 0.04$ \\
\hline\hline
\end{tabular}
\end{center}
\caption{The average modified mass and vector self-energies of the doubly heavy  $\Xi^{(')}_{QQ'}$ and $\Omega^{(')}_{QQ'}$ baryons in GeV at the saturation nuclear matter density together with their vacuum mass values.}
\label{tab3}
\end{table}
\begin{widetext} 
\begin{table*}[t]
	\addtolength{\tabcolsep}{10pt}
	\begin{center}\resizebox{\textwidth}{!}{\begin{tabular}{lccccccccc}\hline \hline 
 $m(\rho=0)$ & PS  &\cite{ALIEV201259} & \cite{Wang2018} & \cite{PhysRevD.78.094007}  &  \cite{Shah2017}/ \cite{Shah2016} & \cite{PhysRevD.96.114006} & \cite{Yu:2018com}  \\
\hline\hline
$\Xi_{cc}$  & $3.65 \pm 0.05$ & $3.72(0.20)$ & $3.63^{+0.08}_{-0.07}$ & $4.26\pm0.19$ & $3.511$ & $3.606$ & $3.63\pm0.02$ \\ 
$\Omega_{cc}$   & $3.66 \pm 0.09$ & $3.73(0.20)$ & $3.75^{+0.08}_{-0.09}$ & $4.25\pm0.20$ & $3.650$ & $3.715$ & $3.73\pm0.02$ \\
$\Xi_{bc}$  & $6.49 \pm 0.04$ & $6.72(0.20)$ & - & $6.75\pm0.05$ & $6.914$ & - & $6.99\pm0.02$\\
$\Omega_{bc}$   & $6.51 \pm 0.05$ & $6.75(0.30)$ & - & $7.02\pm0.08$ & $7.136$ & - & $7.09\pm0.01$ \\
$\Xi_{bb}$  & $9.17 \pm 0.06$  & $9.96(0.90)$ & $10.22^{+0.07}_{-0.07}$ & $9.78\pm0.07$ & $10.312$ & $10.138$ & $10.31\pm0.01$\\
$\Omega_{bb}$   &  $9.20 \pm 0.06$ & $9.97(0.90)$ & $10.33^{+0.07}_{-0.08}$ & $9.85\pm0.07$ & $10.446$ & $10.230$ & $10.37\pm0.01$ \\
$\Xi'_{bc}$  & $6.53 \pm 0.09$ & $6.79(0.20)$ & - & $6.95\pm0.08$ & - & - & $7.01\pm0.02$\\
$\Omega'_{bc}$   &$7.01 \pm 0.04$ & $6.80(0.30)$ & - & $7.02\pm0.08$ & - & - & $7.10\pm0.01$ \\
\hline\hline
 $m(\rho=0)$ & \cite{PhysRevD.83.056006} & \cite{PhysRevD.70.094004} & \cite{PhysRevD.66.034030} & \cite{PhysRevD.90.094007} & \cite{PhysRevD.52.1722} & \cite{PhysRevD.90.094507} & \cite{Migura2006} \\
\hline\hline
$\Xi_{cc}$  & $3.52\sim3.56$ & 3.55 & $3.48$ & $3.627 \pm 0.012$ & $3.660\pm 0.07$ & $3.610$ & $3.642$ \\
$\Omega_{cc}$ & $3.62\sim3.65$ & 3.73  & $3.59$ & $-$ & $3.740\pm0.08$ & $3.738$ & $3.732$ & - \\
$\Xi_{bc}$  & $6.83\sim6.85$ & 6.80 & $6.82$ & $6.914 \pm 0.013$ & $6.990\pm 0.09$ & $6.943$ & - \\
$\Omega_{bc}$   & $6.94\sim6.95$ & 6.98 & $6.93$ & $-$ & $7.060\pm0.09$ & 6.998& -\\
$\Xi_{bb}$  & $10.08\sim10.10$ & 10.10 & $10.09$ & $10.162 \pm 0.012$ & $10.340 \pm 0.10$ & $10.143$ & -\\
$\Omega_{bb}$   & $10.18\sim10.19$  & 10.28 & $10.21$ & $-$ & $10.370\pm0.10$  & $10.273$ &- \\
$\Xi'_{bc}$  & - & 6.87 & $6.85$ & $6.933 \pm 0.012$ & $7.040 \pm 0.09$ & $6.959$  & -\\
$\Omega'_{bc}$  & - & 7.05 & $6.97$ & $-$ & $7.090\pm0.09$ & $7.032$ &- \\
\hline\hline
\end{tabular}}
\end{center}
\caption{Vacuum masses of the doubly heavy  $\Xi^{(')}_{QQ'}$ and $\Omega^{(')}_{QQ'}$ baryons in GeV compared to other theoretical  predictions. PS means present study.}\label{tab4}
\end{table*}
\end{widetext}
At the end of this section, we would like to compare the vacuum mass values of  the spin$ -1/2 $ doubly heavy  $\Xi^{(')}_{QQ'}$ and $\Omega^{(')}_{QQ'}$ baryons obtained from the derived sum rules in  the $  \rho\rightarrow 0  $ limit with the theoretical predictions as well as the existing experimental data in $\Xi_{cc}$ channel. Table \ref{tab4} is presented in this respect.   As seen from this table, the results obtained by using different approaches are over all consistent/close with/to each other within the errors.  There are some channels that some predictions show considerable differences with other predictions. For instance in $\Xi_{cc}$ channel, the results of  Refs. \cite{PhysRevD.78.094007} and \cite{PhysRevD.66.034030} differ from the other predictions, considerably. The former has a prediction a bit larger and the later has the one a bit smaller than the other theoretical results. Our prediction on the mass of $\Xi_{cc}$,  $3.65 \pm 0.05$, is in a nice agreement with the experimental result of LHCb collaboration,  $3621.40\pm72$ (stat.) $\pm ~0.27$ (syst.) $\pm~14(\Lambda_c^+)$ MeV/$c^2$ \cite{PhysRevLett.119.112001}. Our predictions on the mass of other members together with the predictions of other theoretical models can shed light on the future experiments aiming to hunt the doubly heavy baryons and measure their properties. 

\section{Concluding Remarks}
After the discovery of the $\Xi_{cc}^{++}$ state, as a member of the doubly heavy  spin$ -1/2 $ baryon's family,  by LHCb collaboration in 2017 and the tension between the LHCb result with the previous SELEX data has put the subject of doubly heavy baryons at  the center of interests in hadron physics. With the developments in experimental side, it is expected that other members of the family will be discovered in  near future. Naturally, many theoretical studies try to report their predictions on the parameters of the doubly heavy baryons using variety of models and approaches. The studies are mainly done in the vacuum. The present study is the first comprehensive work discussing these baryons both in vacuum and medium with finite density. Thus, in the present work, we derive the masses and vector self energies of the doubly heavy baryons with both the symmetric and anti-symmetric currents in terms of the QCD degrees of freedom, density of the medium, in-medium non-perturbative operators of different dimensions as well as  continuum threshold, Borel mass parameter and mixing parameter $\beta$ as the helping parameters entering the calculations. With the standard prescriptions of the in-medium QCD sum rules method we restricted the auxiliary parameters to find their reliable working window. We observed that the medium enlarges the working window of the mixing parameter  $\beta$, considerably. Using the reliable working intervals of the helping parameters we extracted the masses and vector self-energies of the baryons under consideration  at saturated nuclear matter  density. It is observed that the $\Omega^{(')}_{QQ'}$ baryons do not overall  see the medium at all, while the  parameters of  $\Xi^{(')}_{QQ'}$ baryons are affected by the medium considerably.  Such that, at saturated nuclear matter density, the masses of the  baryons  $\Xi_{cc}, \Xi_{bc}$ and $\Xi_{bb}$ reach to  $0.81$,  $0.79$  and $0.77$ of their vacuum values, respectively. The negative shifts on the masses due to the medium show that these baryons are attracted (scalar self-energy attraction) by the medium.  The $\Omega^{(')}_{QQ'}$ baryons gain small vector self-energies at saturated nuclear matter density. 
The positive and  large vector self-energies of the $\Xi^{(')}_{QQ'}$ baryons indicate that these baryons endure strong vector repulsion from the medium. 

We investigated the behavior of the   $m^*/m$ with respect to $\rho/\rho_{sat}$ for the doubly heavy spin$ -1/2 $ baryons  in the range $ \rho\in [0, 1.4] \rho_{sat}$. We observed that the masses of  the doubly heavy $\Omega'_{QQ'}$ baryons remain unchanged even up to $\rho= 1.4 \rho_{sat}$.  While  the doubly heavy baryons $\Xi^{(')}_{QQ'}$   are affected by the medium, considerably. Such that the masses of the  baryons  $\Xi_{cc}, \Xi_{bc}$ and $\Xi_{bb}$ reach to $42\%$, $40\%$ and $24\%$ of their vacuum values at the end point, respectively.

We extracted the masses of all members  in $\rho\rightarrow0$ limit as well and compared the results with other theoretical vacuum predictions. Our prediction on the mass of $\Xi_{cc}$ is in a nice agreement with the experimental data of LHCb collaboration \cite{PhysRevLett.119.112001}. Our predictions on the vacuum masses of other members may help experimental groups in the search for these baryons,  which are natural outcomes of the quark model. Our results on the in-medium masses and vector-self energies of the doubly heavy baryons may shed light on the future in-medium experiments and help physicists in analyzing the results of such experiments. \\\\

\begin{widetext}
 \section*{Appendix A: THE LIGHT AND HEAVY QUARKS PROPAGATORS AND THEIR IN-MEDIUM INGREDIENTS}
In this Appendix, we present the explicit expressions of the in-medium quarks propagators including their ingredients: the  in-medium quark, gluon and mixed condensates.  In the  calculations, the light quark propagator is  used in  the fixed point gauge,

\begin{eqnarray}\label{lightq}
S_q^{ij}(x)&=&
\frac{i}{2\pi^2}\delta^{ij}\frac{1}{(x^2)^2}\not\!x
-\frac{m_q }{ 4\pi^2} \delta^ { ij } \frac { 1}{x^2} + \chi^i_q(x)\bar{\chi}^j_q(0) 
-\frac{ig_s}{32\pi^2}F_{\mu\nu}^A(0)t^{ij,A}\frac{1}{x^2}[\not\!x\sigma^{\mu\nu}+\sigma^{\mu\nu}\not\!x] +\cdots, \nonumber\\
\end{eqnarray}

where  $\chi^i_q$ and $\bar{\chi}^j_q$ are the Grassmann background quark fields, $F_{\mu\nu}^A$ are classical background gluon fields, and $t^{ij,A}=\frac{\lambda ^{ij,A}}{2}$ with $
\lambda ^{ij, A}$ being  the standard Gell-Mann matrices. The heavy quark propagator is given as

\begin{eqnarray}\label{heavyQ}
S_Q^{ij}(x)&=&\frac{i}{(2\pi)^4}\int d^4k e^{-ik \cdot x} \left\{\frac{\delta_{ij}}{\!\not\!{k}-m_Q}
-\frac{g_sF_{\mu\nu}^A(0)t^{ij,A}}{4}\frac{\sigma_{\mu\nu}(\!\not\!{k}+m_Q)+(\!\not\!{k}+m_Q)
\sigma_{\mu\nu}}{(k^2-m_Q^2)^2}\right.\nonumber\\
&&\left.+\frac{\pi^2}{3} \langle \frac{\alpha_sGG}{\pi}\rangle\delta_{ij}m_Q \frac{k^2+m_Q\!\not\!{k}}{(k^2-m_Q^2)^4}+\cdots\right\} \, .
\end{eqnarray}

By replacing these explicit forms of the light and heavy quark propagators in the correlation function in Eqs.~(\ref{symmet}-\ref{asymmet}), the products
of the Grassmann background quark fields and classical background gluon fields, which correspond to the ground-state matrix elements of the corresponding quark and gluon operators \cite{COHEN1995221} are obtained,
\begin{eqnarray}\label{fields}
\chi_{a\alpha}^{q}(x)\bar{\chi}_{b\beta}^{q}(0)&=&\langle q_{a\alpha}(x)\bar{q}_{ b\beta}(0)\rangle_{\rho},
 \nonumber \\
F_{\kappa\lambda}^{A}F_{\mu\nu}^{B}&=&\langle
G_{\kappa\lambda}^{A}G_{\mu\nu}^{B}\rangle_{\rho}, \nonumber \\
\chi_{a\alpha}^{q}\bar{\chi}_{b\beta}^{q}F_{\mu\nu}^{A}&=&\langle
q_{a\alpha}\bar{q}_{ b\beta}G_{\mu\nu}^{A}\rangle_{\rho},
\end{eqnarray}
where, $\rho$ is the medium density. The matrix elements in the right hand sides of equations in Eq.~(\ref{fields}) contain the in-medium quark, gluon and mixed condensates, whose explicit forms   are given as \cite{COHEN1995221}: 

1-) Quark condensate:

\begin{eqnarray} \label{quarkfield}
&& \langle q_{a\alpha}(x)\bar{q}_{b\beta}(0)\rangle_{\rho}=-\frac{\delta_{ab}}{12} \Bigg[\Bigg(\langle\bar{q}q\rangle_{\rho}+x^{\mu}\langle\bar{q}D_{\mu}q\rangle_{
\rho} + \frac{1}{2}x^{\mu}x^{\nu}\langle\bar{q}D_{\mu}D_{\nu}q\rangle_{\rho} +...\Bigg)\delta_{\alpha\beta}
 \nonumber \\
&&+\Bigg(\langle\bar{q}\gamma_{\lambda}q\rangle_{\rho}+x^{\mu}\langle\bar{q}
\gamma_{\lambda}D_{\mu} q\rangle_{\rho}+\frac{1}{2}x^{\mu}x^{\nu}\langle\bar{q}\gamma_{\lambda}D_{\mu}D_{\nu}
q\rangle_{\rho}
+...\Bigg)\gamma^{\lambda}_{\alpha\beta} \Bigg],\nonumber \\
\end{eqnarray}
2-) Gluon condensate:
\begin{eqnarray}\label{gluonfield}
 \langle G_{\kappa\lambda}^{A}G_{\mu\nu}^{B}\rangle_{\rho}&=&\frac{\delta^{AB}}{96}
\Bigg[ \langle G^{2}\rangle_{\rho}(g_{\kappa\mu}g_{\lambda\nu}-g_{\kappa\nu}g_{\lambda\mu})+O(\langle \textbf{E}^{2}+\textbf{B}^{2}\rangle_{\rho})\Bigg],
\end{eqnarray}
where the term $O(\langle \textbf{E}^{2}+\textbf{B}^{2}\rangle_{\rho})$ is neglected because of its small contribution. 

3-) Quark-gluon mixed condensate:
\begin{eqnarray} \label{mixedfield}
&&\langle g_{s}q_{a\alpha}\bar{q}_{b\beta}G_{\mu\nu}^{A}\rangle_{\rho}=-\frac{t_{ab}^{A }}{96}\Bigg\{\langle g_{s}\bar{q}\sigma\cdot Gq\rangle_{\rho}
\Bigg[\sigma_{\mu\nu}+i(u_{\mu}\gamma_{\nu}-u_{\nu}\gamma_{\mu}) \!\not\! {u}\Bigg]_{\alpha\beta} +\langle g_{s}\bar{q}\!\not\! {u}\sigma\cdot Gq\rangle_{\rho} \Bigg[\sigma_{\mu\nu}\!\not\!
{u} \nonumber \\
&&+i(u_{\mu}\gamma_{\nu}-u_{\nu}\gamma_{\mu} )\Bigg]_{\alpha\beta}-4\Bigg(\langle\bar{q}u\cdot D u\cdot D
q\rangle_{\rho}
+im_{q}\langle\bar{q}
\!\not\! {u}u\cdot D q\rangle_{\rho}\Bigg) \Bigg[\sigma_{\mu\nu}+2i(u_{\mu}\gamma_{\nu}-u_{\nu}\gamma_{\mu}
)\!\not\! {u}\Bigg]_{\alpha\beta}\Bigg\},
\nonumber \\
\end{eqnarray}
where $D_\mu=\frac{1}{2}(\gamma_\mu \Dslash+\Dslash\gamma_\mu)$. The modified in-medium different condensates in Eqs.~(\ref{quarkfield}-\ref{mixedfield}) are  presented as:
\begin{eqnarray} \label{ner}
\langle\bar{q}\gamma_{\mu}q\rangle_{\rho}&=&\langle\bar{q}\!\not\!{u}q\rangle_{\rho} u_{\mu} ,\nonumber \\
\langle\bar{q}D_{\mu}q\rangle_{\rho}&=&\langle\bar{q}u\cdot D q\rangle_{\rho} u_{\mu}=-im_{q}\langle\bar{q}\!\not\!{u}q\rangle_{\rho} u_{\mu}  ,\nonumber \\
\langle\bar{q}\gamma_{\mu}D_{\nu}q\rangle_{\rho}&=&\frac{4}{3}\langle\bar{q} \!\not\! {u}u\cdot D
q\rangle_{\rho}(u_{\mu}u_{\nu}-\frac{1}{4}g_{\mu\nu}) +\frac{i}{3}m_{q}
\langle\bar{q}q\rangle_{\rho}(u_{\mu}u_{\nu}-g_{\mu\nu}),
\nonumber \\
\langle\bar{q}D_{\mu}D_{\nu}q\rangle_{\rho}&=&\frac{4}{3}\langle\bar{q}
u\cdot D u\cdot D
q\rangle_{\rho}(u_{\mu}u_{\nu}-\frac{1}{4}g_{\mu\nu}) -\frac{1}{6} \langle
g_{s}\bar{q}\sigma\cdot Gq\rangle_{\rho}(u_{\mu}u_{\nu}-g_{\mu\nu}) , \nonumber \\
\langle\bar{q}\gamma_{\lambda}D_{\mu}D_{\nu}q\rangle_{\rho}&=&2\langle\bar{q}
\!\not\! {u}u\cdot D u\cdot D q\rangle_{\rho}
\Bigg[u_{\lambda}u_{\mu}u_{\nu} -\frac{1}{6}
(u_{\lambda}g_{\mu\nu}+u_{\mu}g_{\lambda\nu}+u_{\nu}g_{\lambda\mu})\Bigg]\nonumber\\
&&-\frac{1}{6} \langle g_{s}\bar{q}\!\not\! {u}\sigma\cdot
Gq\rangle_{\rho}(u_{\lambda}u_{\mu}u_{\nu}-u_{\lambda}g_{\mu\nu}).\nonumber
\\
\end{eqnarray}

\section*{Appendix B: NUMERICAL INPUTS}
In numerical calculations, the vacuum condensates are used at a renormalization scale of $1$ GeV : $\rho_{sat}=0.11^3 ~$GeV$^3$, $ \langle q^{\dag} q\rangle_{\rho}=\frac{3}{2}\rho$,  $ \langle s^{\dag} s\rangle_{\rho}=0$ \cite{COHEN1995221}, $ \langle \bar{q}q\rangle_{0}=(-0.241)^3$~GeV$^3$ \cite{IOFFE2006232},  $\langle \bar{s}s\rangle_{0} = 0.8 \langle \bar{q}q\rangle_{0}$  \cite{COHEN1995221},  $\langle \bar{q}q\rangle_{\rho}=\langle \bar{q}q\rangle_{0}+\frac{\sigma_{\pi N}}{2 m_q}\rho$ \cite{PhysRevC.47.2882}, $m_q=\frac{m_u+m_d}{2}=0.00345$~GeV \cite{PhysRevD.98.030001}, $\langle \bar{s}s\rangle_{\rho}=\langle \bar{s}s\rangle_{0}+y\frac{\sigma_{\pi N}}{2 m_q}\rho$, $y=0.05\pm0.01$ (the average of  values obtained in Refs. \cite{PhysRevD.87.074503} and \cite{Dinter:2011za}), $\langle \frac{\alpha_s}{\pi}G^2\rangle_{0}=(0.33\pm0.04)^4~$GeV$^4$, $\langle \frac{\alpha_s}{\pi}G^2\rangle_{\rho}=\langle \frac{\alpha_s}{\pi}G^2\rangle_{0}-(0.65\pm 0.15)~$GeV$~\rho$, $\langle q^{\dag}iD_0 q\rangle_{\rho}=0.18~$GeV$~\rho$, $\langle s^{\dag}iD_0 s\rangle_{\rho}=\frac{m_s\langle \bar{s}s\rangle_{\rho}}{4}+0.02~$GeV$~\rho$, $\langle \bar{q}iD_0q\rangle_{\rho}=\langle \bar{s}iD_0s\rangle_{\rho}=0$, $\langle \bar{q}g_s\sigma G q\rangle_{0}=m_0^2 \langle \bar{q}q\rangle_{0}$, $\langle \bar{s}g_s\sigma G s\rangle_{0}=m_0^2 ~\langle \bar{s}s\rangle_{0}$  \cite{COHEN1995221}, $m_0^2=0.8 ~$GeV$^2$ \cite{IOFFE2006232}, $\langle \bar{q}g_s\sigma G q\rangle_{\rho}=\langle \bar{q}g_s\sigma G q\rangle_{0}+ 3~GeV^2~\rho$, $\langle \bar{s}g_s\sigma G s\rangle_{\rho}=\langle \bar{s}g_s\sigma G s\rangle_{0}+ 3y~$GeV$^2~\rho$, $\langle q^{\dag}g_s\sigma G q\rangle_{\rho}=-0.33 ~GeV^2~\rho$, $\langle q^{\dag}iD_0 iD_0 q\rangle_{\rho}=0.031~$GeV$^2~\rho-\frac{1}{12}\langle q^{\dag}g_s\sigma G q\rangle_{\rho}$, $\langle s^{\dag}g_s\sigma G s\rangle_{\rho}=-0.33y ~$GeV$^2~\rho$ and $\langle s^{\dag}iD_0 iD_0 s\rangle_{\rho}=0.031y~$GeV$^2~\rho-\frac{1}{12}\langle s^{\dag}g_s\sigma G s\rangle_{\rho}$  \cite{COHEN1995221,PhysRevC.47.2882}.  For the pion nucleon sigma term we use $\sigma_{\pi N}=0.059$~GeV \cite{PhysRevD.85.051503}. 

The light quark masses are used at a renormalization scale $1$ GeV, as well: $m_u=2.16_{-0.26}^{+0.49}$~MeV, $m_d=4.67_{-0.17}^{+0.48}$~MeV, $m_s=93_{-5}^{+11}$~MeV \cite{PhysRevD.98.030001}. For the heavy quarks, we use the pole masses. The relation between the pole mass $m_Q$ and $\overline{\textrm{MS}}$ mass $\overline{m}_Q$ for the heavy quarks  in three loops  is given as \cite{Gray1990,Broadhurst1991,PhysRevLett.83.4001,MELNIKOV200099}
\begin{eqnarray}\frac{}{}
m_Q & = &\overline{m}_Q(\overline{m}_Q)\Bigg\{1+\frac{4\overline{\alpha}_s(\overline{m}_Q)}{3\pi} +  \Bigg[-1.0414 \Sigma_k\Big(1-\frac{4}{3}\frac{\overline{m}_{Q_k}}{\overline{m}_Q}\Big)+ 13.4434\Bigg]\Big[\frac{\overline{\alpha}_s(\overline{m}_Q)}{\pi}\Big]^2\nonumber \\
 &+&\Bigg[0.6527N_L^2 -26.655N_L+190.595\Bigg]\Big[\frac{\overline{\alpha}_s(\overline{m}_Q)}{\pi}\Big]^3\Bigg\}
\end{eqnarray}
where $\overline{\alpha}_s(\mu)$ is the strong interaction coupling constant in the $\overline{\textrm{MS}}$ scheme, and the sum
over $k$ extends over the $N_L$ flavors $Q_k$ lighter than $Q$. Using the $\overline{\textrm{MS}}$ mass values presented in PDG, one gets  $m_b=4.78\pm0.06$~GeV and  $m_c= 1.67\pm 0.07$ GeV for the bottom and charm pole masses, which are used in numerical calculations.

 Note that, at dense medium, each condensate  is expanded up to the first order in nucleon density as $\langle\hat{O}\rangle_{\rho}=\langle\hat{O}\rangle_{0} +\langle\hat{O}\rangle_{N}\rho$, where $\langle\hat{O}\rangle_{0}$ is the vacuum expectation value of the operator $  \hat{O}$ and $\langle\hat{O}\rangle_{N}$ is its expectation  value between one-nucleon states \cite{COHEN1995221,PhysRevC.47.2882,PhysRevD.98.030001}.

\bibliography{refs.bib}

\end{widetext}
%

\end{document}